\documentclass[prb,twocolumn,floatfix,a4paper,superscriptaddress,nobalancelastpage]{revtex4}
\usepackage{amsmath,amsfonts,amssymb,graphics,graphicx,epsfig,color,bbm}
\usepackage{times}

\begin{document}

\title{Polar bosons in one-dimensional disordered optical lattices}
\author{Xiaolong Deng}
\affiliation{Institut f\"ur Theoretische Physik, Leibniz Universit\"at Hannover, Appelstr. 2, D-30167 Hannover, Germany}
\author{Roberta Citro}
\affiliation{Dipartimento di Fisica ``E. R. Caianiello'' and Spin-CNR, Universit\`a degli Studi di Salerno, Salerno, Italy}
\author{Edmond Orignac}
\affiliation{Laboratoire de Physique de l'\'Ecole Normale Sup\'erieure de Lyon, CNRS-UMR5672, 69364 Lyon Cedex 7, France}
\author{Anna Minguzzi}
\affiliation{Universit\'e Grenoble I and CNRS, Laboratoire de Physique et Mod\'elisation,
des Milieux Condens\'es UMR 5493, Maison des Magist\`eres, B.P. 166, 38042 Grenoble, France}
\author{Luis Santos}
\affiliation{Institut f\"ur Theoretische Physik, Leibniz Universit\"at Hannover, Appelstr. 2, D-30167 Hannover, Germany}
\begin{abstract}
We analyze the effects of disorder and quasi-disorder on the ground-state properties of ultra-cold polar bosons in optical lattices. 
We show that the interplay between disorder and inter-site interactions leads to rich phase diagrams. 
A uniform disorder leads to a Haldane-insulator phase with finite parity order, whereas the density-wave phase becomes 
a Bose-glass at very weak disorder. For quasi-disorder, the Haldane insulator connects with a gapped generalized 
incommesurate density wave without an intermediate critical region.
\end{abstract}

\date{\today}

\maketitle

The interplay between disorder and interactions plays a crucial role in the physics of strongly-correlated systems~\cite{Lee1985}.  
Disorder in non-interacting systems leads to Anderson localization~\cite{Anderson1958}, which in one dimension 
occurs for vanishingly small disorder~\cite{Abrahams1979}. For the particular case of bosons in a lattice potential, interactions 
have been shown, both in 1D~\cite{Giamarchi1987,Giamarchi1988,Prokofev1998,Rapsch1999} and higher dimensions~\cite{Fisher1989,Guraire2009}, to induce in the presence of disorder a phase diagram characterized by three phases: a superfluid~(SF) phase, a gapless localized incompressible phase known as Bose-Glass~(BG), and a Mott-insulator~(MI) occuring at commesurate lattice fillings. 

Ultra-cold atoms in optical lattices offer an extraordinarily controllable scenario for the detailed analysis of the competition between disorder and interactions. Disorder in the on-site energies may be implemented in various ways in these systems, including the use of speckle~\cite{Clement2005,Fort2005,Schulte2005,Chen2008}, binary mixtures~\cite{Vignolo2003,Paredes2005,Gavish2005,Ospelkaus2006}, and bichromatic combinations of two mutually incommensurate lattices~\cite{Fallani2007}. Recently, localization has been experimentally observed in non-interacting cold gases in 1D and 3D speckle~\cite{Billy2008,Kondov2011,Jendrzejewski2012}, and bichromatic potentials~\cite{Roati2008}. Bichromatic lattices constitute a peculiar type of disorder, rather a quasi-disorder, 
realizing the so-called Aubry-Andr\'e model~\cite{Aubry1980}. The effects of interactions in 1D lattices with quasi-disorder have been recently studied~\cite{Roscilde2008,Roux2008,Deng2008,Quasiperiodic-Bond}. Particularly interesting 
is the existence of a gapped localized phase, the so-called incommensurate density wave~(ICDW), which results from the quasi-periodicity of the potential. 

Polar gases are attracting a growing attention mostly motivated by experiments on atoms with large magnetic moments
~\cite{Griesmaier2005,Lu2011,Aikawa2012}, and especially by recent groundbreaking experiments on polar molecules~\cite{Ospelkaus2010}. Due to the dipole-dipole interaction, these gases present an exceedingly rich 
physics~\cite{Baranov2008,Lahaye2009}. Polar lattice gases are particularly interesting~\cite{Trefzger2011}.
In particular, intersite interactions 
may allow for the realization of the so-called Haldane-insulator~(HI) phase~\cite{DallaTorre2006}, a gapped phase characterized by 
a nonlocal string order parameter.

In this paper we show that the interplay between on-site and inter-site interactions and disorder leads to a rich physics 
for lattice bosons with nearest neighbor interactions. In particular, in the presence of uniform disorder the HI phase 
is preserved, although with finite parity, up to a finite disorder, where a phase transition into a BG is produced. On the contrary, in the presence of quasi-disorder the HI is connected, without any intermediate critical region, to a gapped generalized ICDW phase occurring for non-polar gases. Other phases are discussed in detail for both types of disorder.

As mentioned above, the main qualitatively new feature of polar lattice gases concerns the significant 
inter-site interactions. Polar interactions between sites placed $j>0$ sites apart decay as $1/j^3$. Although 
interactions for $j>1$ do play a role in the physics of polar gases, especially in what concerns the existence of crystalline phases 
at any fractional filling~(Devil's staircase~\cite{Burnell2009,Capogrosso-Sansone2010}) 
for very weak on-site interactions and sufficiently large
dipoles, the most relevant properties of polar lattices gases may be understood 
from a model with only nearest-neighbor interactions:
\begin{eqnarray}
H &=& -t\sum_{i} (b^{\dagger}_i b_{i+1}+H.c.) + \frac{U}{2}\sum^N_{i=1} n_i(n_i -1) \nonumber\\
& & + V\sum_{i} n_in_{i+1} + \sum_i \epsilon_i n_i,
\end{eqnarray}
with $b^{\dagger}_i$, $b_i$ the creation/annihilation operators for bosons at site $i$, $n_i=b^{\dagger}_ib_i$, 
$t$ the hopping amplitude, $U$~($V$) the on-site~(inter-site) interaction, and $\epsilon_i$ the on-site energy.

\begin{figure}[t]
\resizebox{2.2in}{!}{\includegraphics{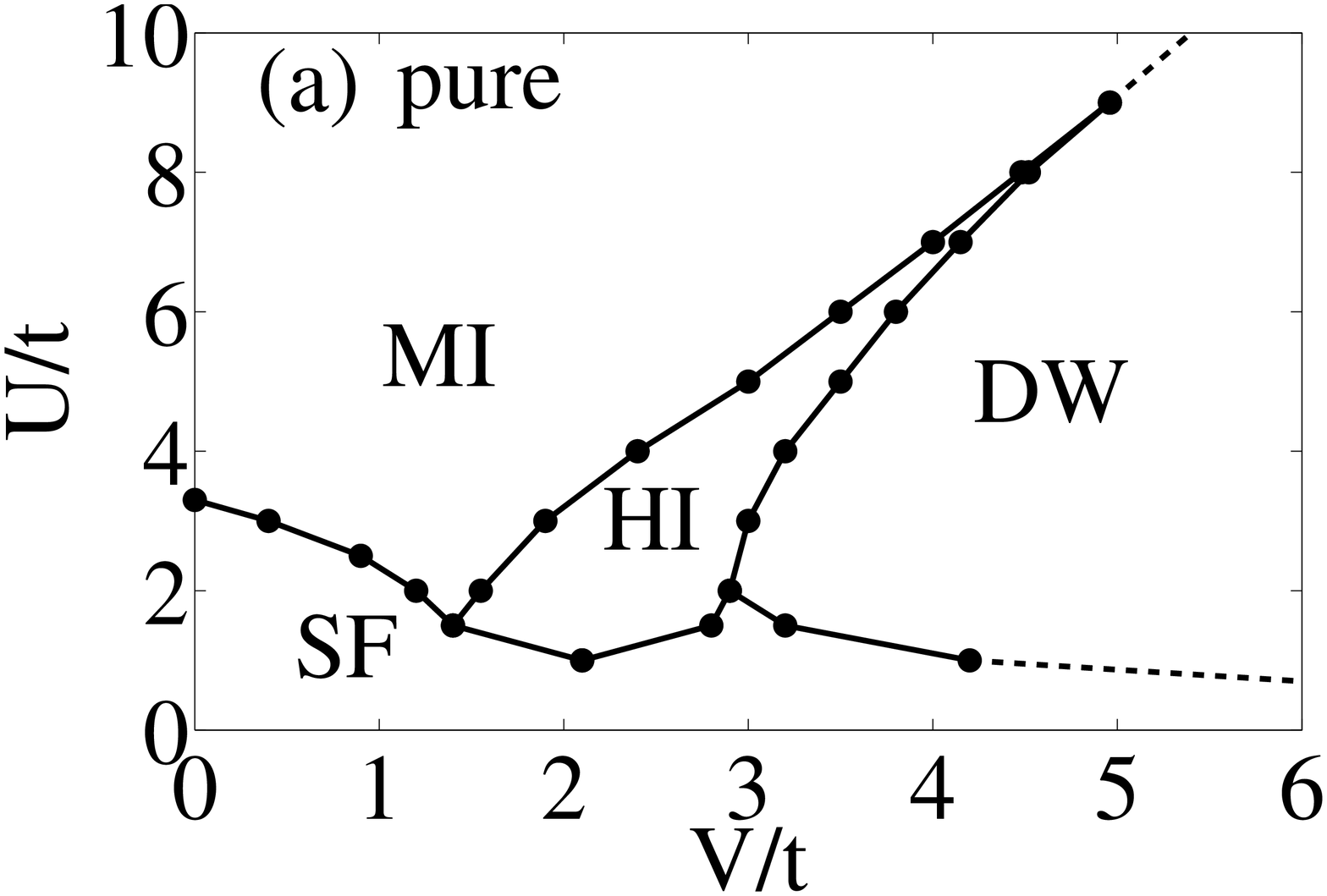}} 
\resizebox{2.2in}{!}{\includegraphics{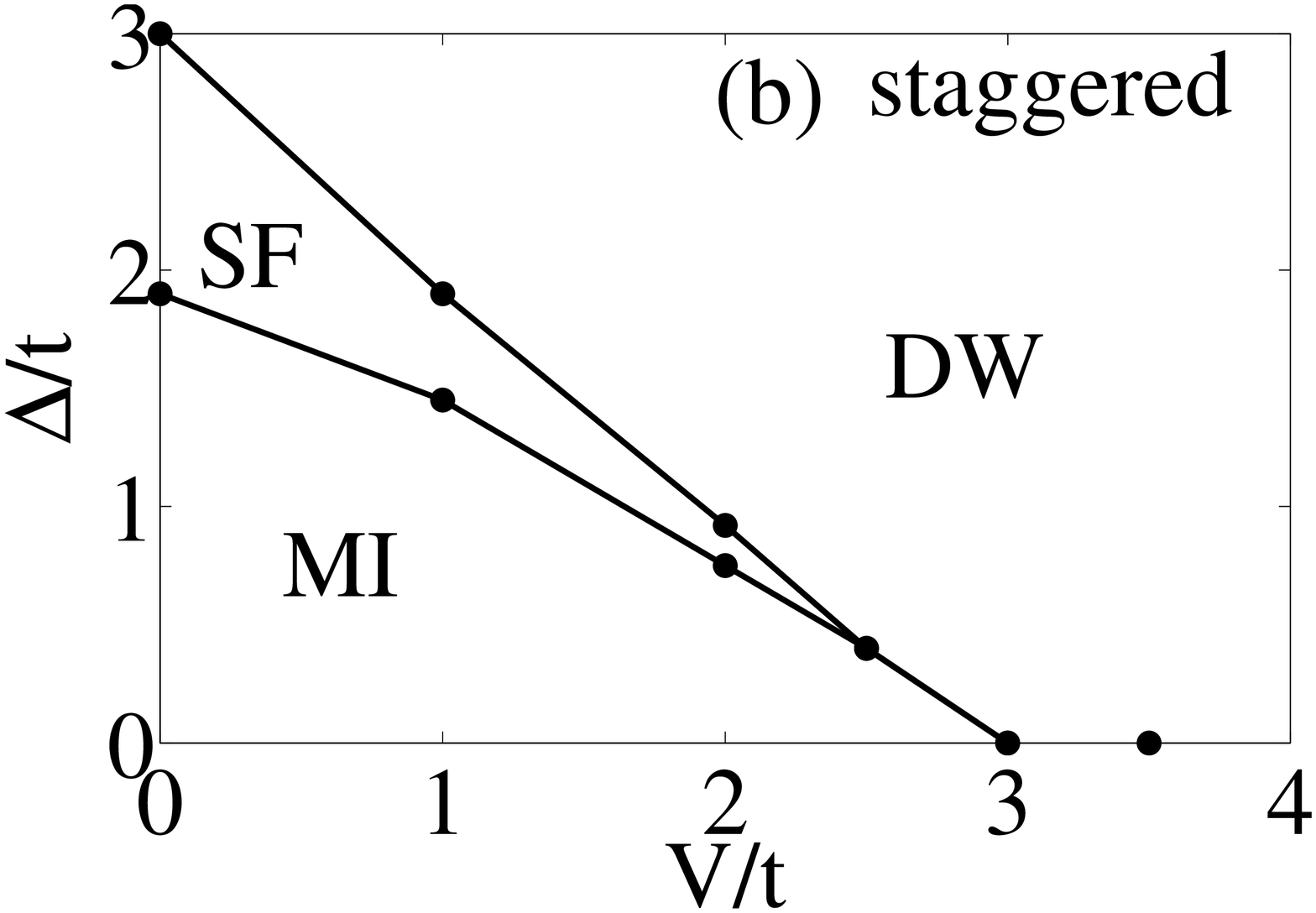}}
\resizebox{2.2in}{!}{\includegraphics{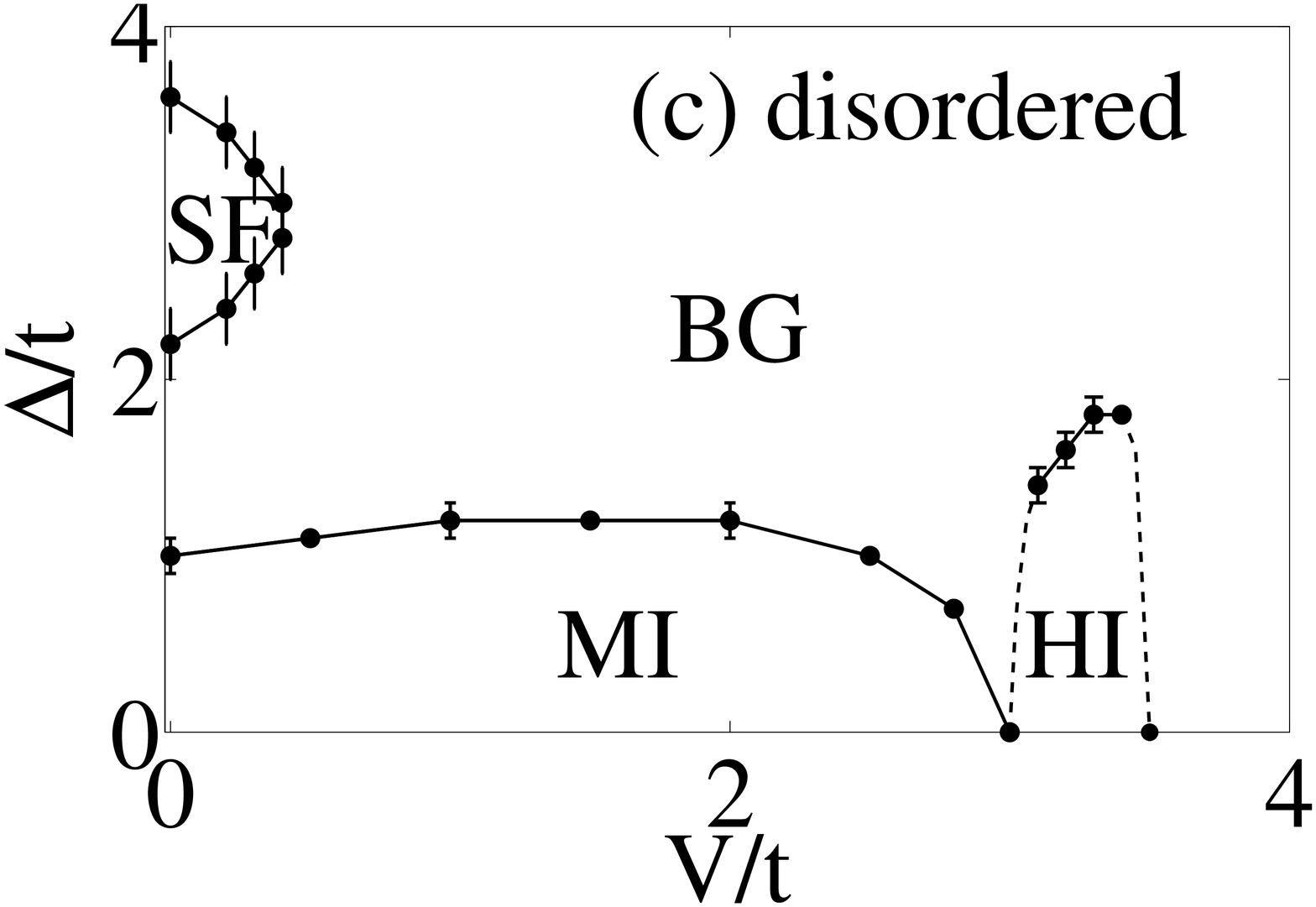}}
\resizebox{2.2in}{!}{\includegraphics{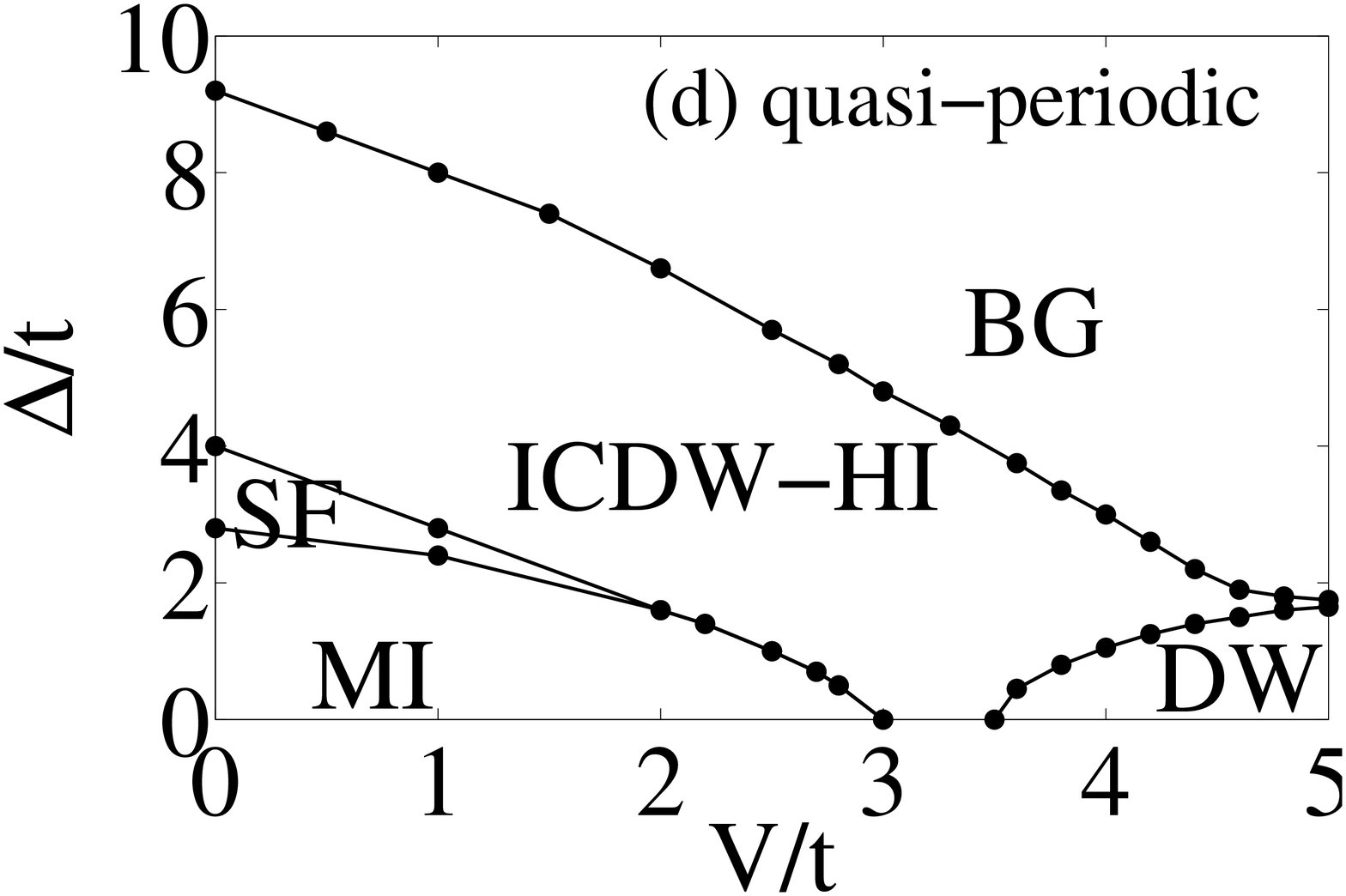}}
\vspace*{-0.2cm}
\caption{Phase diagrams for 
(a) unperturbed case; (b) staggered on-site energy; (c) uniform disorder; (d)  
quasi-disorder. Figures (b)--(d) are obtained for $U/t=5$. See text for details.}
\label{fig:1}
\end{figure}

In the following we consider an average filling factor $n=1$. Assuming small deviations from this average, we may introduce an effective 
spin via the standard Holstein-Primakoff~(HP) transformation $S_i^z=\delta n_i=1-n_i$, $S_i^+=\sqrt{2-n_i}b_i$. The resulting 
spin Hamiltonian resembles to a large extent an antiferromagnetic spin-$1$ XXZ model 
with uniaxial single-ion anisotropy (the mapping is however 
imperfect, especially at low $U$, due to occupations $n>2$, and extra terms appearing when introducing the HP 
transformation):
\begin{equation}
H=J\sum_i \! \left [S_i^xS_{i+1}^x\! + \! S_i^yS_{i+1}^y \! +\! \Delta S_i^zS_{i+1}^z\! +\! D(S_i^z)^2 \right ],\label{eq:H-Spin}
\end{equation}
with $J=2t$, $D=U/4t$, and $\Delta=V/2t$. For $D>0$ and $\Delta>0$ three ground-state phases are possible~\cite{Chen2003}, a N\'eel antiferromagnet~(AF) for sufficiently large $\Delta>0$, a large-$D$ phase for dominant single-ion anisotropy, and the gapped Haldane phase~\cite{Haldane1983} characterized by a non-local string-order~\cite{DenNijs1989}.

Unraveling the HP transformation, the equivalent bosonic phases are: a density 
wave~(DW) phase, characterized by a finite ${\cal O}_{DW}=\lim_{|i-j|\rightarrow\infty}\langle (-1)^{i-j}\delta n_i \delta n_j\rangle$, 
and a static structure factor $S(k)=\frac{1}{N^2}\sum_{i,j} e^{ik(i-j)}\langle n_i n_j\rangle$~(with $N$ the number of sites) 
peaked at $k=\pi$; a MI with hidden parity order ${\cal O}_{P}=\lim_{|i-j|\rightarrow\infty}\langle (-1)^{\sum_{i<l<j}\delta n_l}\rangle$~\cite{Berg2008}~(recently measured in site-resolved experiments~\cite{Endres2011}); and a 
Haldane-insulator~(HI) phase, with finite string-order ${\cal O}_{S}=\lim_{|i-j|\rightarrow\infty}\langle \delta n_i (-1)^{\sum_{i<l<j}\delta n_l}\delta n_j\rangle$~\cite{Berg2008,DallaTorre2006}.
All the previous phases are insulating, being characterized by an exponentially decaying single-particle correlation function 
$G(r)=\langle b_i^\dag b_{i+r}\rangle /\sqrt{\langle n_i\rangle\langle n_{i+r} \rangle}$~\cite{Roux2008}. 

The establishment of these orders has been analysed numerically using density-matrix renormalization group~(DMRG~\cite{White1992}). The phase boundaries are inferred by the study of the single particle correlation function $G(r)$ and its exponent, as explained below. The resulting phase diagram for the pure case is shown in Fig.\ref{fig:1}a. At large $U$ a direct first-order MI-DW transition occurs. At intermediate $U$ a HI occurs in a window $V_c^{(1)}(U)<V<V_c^{(2)}(U)$, as first found with DMRG calculations for soft-core polar bosons ~\cite{Berg2008,DallaTorre2006,footnote0}. The  MI-HI Gaussian transition line  is a Luttinger liquid with Luttinger parameter $2>K_c(U)>1/2$~\cite{Berg2008}, and the HI-DW transition line belongs to the Ising universality class \cite{Boschi2003}.  We identify both transition lines  by the algebraic decay of $G(r)$. At low $U$ occupations $n>2$ break down the boson-spin mapping, and a superfluid~(SF) phase may be 
found, which is a Luttinger-liquid characterized by an algebraically 
decaying $G(r)\propto r^{-1/2K}$. A MI to SF Kosterlitz-Thouless~(KT) transition occurs  at sufficiently low $U$ and $V$, and we identify it by searching  when the Luttinger parameter $K>2$~\cite{Berg2008,Giamarchi2004}. 
At low $U$ but large $V$ our DMRG calculations show a DW to SF transition with a critical $K=1/2$~\cite{Giamarchi2004}. 
Below the HI region we find as well a SF region for $K>2$.

We employ below DMRG calculations with open boundary conditions for determining the ground-state phases for 
different types of disorder. Special care must be taken with the edge states in the HI phase, which are polarized by adding one more particle or eliminated by coupling two extra hard-core bosons at the chain edges in order to form a singlet state~\cite{Deng2011}. 
Both procedures destroy the energy gap for the MI and ICDW phases. We hence introduce two different 
definitions for the gap to the first excited state, 
$E_G$ and $E'_G$, which are respectively calculated without and with the mentioned polarization/elimination procedure. 
For the MI and ICDW, $E_G\neq 0$ and $E'_G=0$; for HI and DW by $E_G=0$ and $E'_G\neq 0$; and 
for the gapless phases (SF and BG) $E_G=E'_G=0$. 
The Luttinger exponent $K$ of the SF phases and of the critical lines is extracted by fitting the single-particle correlation function $G(r)$ 
using finite-size conformal corrections~\cite{Cazalilla2004}. The number of particles is conserved and 
the maximal bosonic occupation per site is $n_{max}=8$ for $U \neq 0$~\cite{footnote2}. 
We consider lattice sizes of $N=55$, $89$, $144$ and $233$ sites~(see discussion below). The 
number of optimal states kept in the DMRG ranges from $300$ to $500$. 
Statistical deviations are particularly relevant in the case of uniform disorder discussed below. For that case we have performed 
up to $60$ realizations per case~(we refer to the supplementary material for a discussion on the error bars). 
We have checked that our results converge to the exact values in the non-interacting limit~(where we keep $n_{max}=N$), 
and to earlier calculations for interacting systems with $V=0$~\cite{Prokofev1998}~(see the supplementary material).

As a starting point we first consider the case of a commesurate 
staggered $\epsilon_i=(-1)^i \Delta$, which resembles the case of staggered magnetic fields 
in spin chains. For the latter case it is known~\cite{Tsukano1998a,Tsukano1998b} that the Haldane phase is continuously connected to the AF phase. Due to the explicit symmetry breaking, the AF phase is a singlet and not a doublet, and a Gaussian transition
occurs between large-D and AF phases. We hence expect that the HI phase continuously connect to the DW phase for even vanishingly small $\Delta$, as observed in our numerics~(Fig.~\ref{fig:1}(b)). Note that a finite SF region opens for the bosonic case between the MI and the DW phases, the transition lines being inferred from the decay of $G(r)$ as in the pure case.

We consider next a uniform disorder, in which $\epsilon_i$ has a uniform random distribution in the interval $[-\Delta,\Delta]$. The presence of disorder induces a gapless localized Bose-Glass (BG) phase~\cite{Giamarchi1987,Giamarchi1988,Fisher1989}, separating the MI and  the SF phase. At unit filling, and intermediate $U$ values, for growing $\Delta$ a MI-BG transition, identified by the vanishing of the energy gap,  is followed by a BG-SF transition, and a final SF-BG transition at larger $\Delta$. The SF-BG boundary forms a characteristic finger-like shape~\cite{Prokofev1998,Rapsch1999} and is obtained by searching where SF correlations decay with exponent $K=3/2$~\cite{Giamarchi1987,Giamarchi1988}. 
Our numerics recover these well-known results for $V=0$~(see Fig.~\ref{fig:1}(c) and the supplementary material). 
For growing $V$ the intermediate SF region shrinks and eventually vanishes, whereas the MI phase 
stretches up to a critical $V=V_c^{(1)}$, which marks the MI-HI transition for $\Delta=0$. Close to $V_c^{(1)}$ numerical calculations cannot resolve 
whether a BG opens at the MI-HI boundary for vanishingly small disorder. Although RG calculations to the lowest order predict a stable MI-HI boundary 
at weak disorder\cite{Brunel1998}, RG calculations using $4k_F$ terms to the density~\cite{Orignac1998,Citro-inprep} show that a BG should appear at the MI-HI 
boundary at vanishingly small disorder for $K<3/2$ and at finite disorder for $K>3/2$. For the value $U/t=5$ in the figure, $K=1.18$ and we hence expect a BG 
all the way to $\Delta=0$.
\begin{figure}[t]
\includegraphics[clip=true,width=0.8\columnwidth]{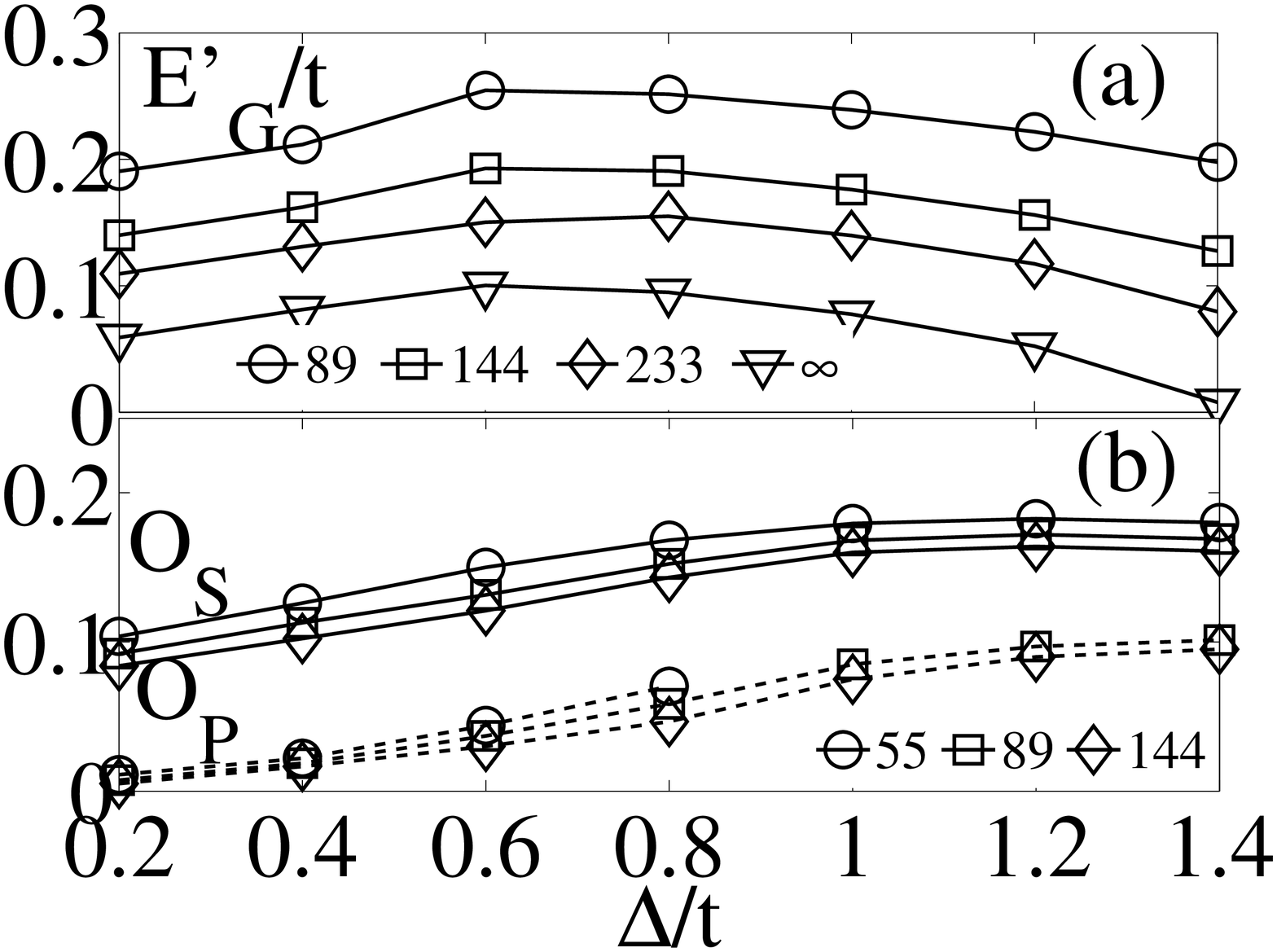}
\vspace*{-0.3cm}
\caption{(a) Gap $E'_G$ as a function of $\Delta$ for uniform disorder 
with $U=5t$, $V=3.1t$, and different lattice sizes; (b) For the same parameters, ${\cal O}_S$ and ${\cal O}_P$, for different lattice sizes. Finite-size scaling is visible.
}
\label{fig:2}
\end{figure}

\begin{figure}[t]
\includegraphics[clip=true,width=0.8\columnwidth]{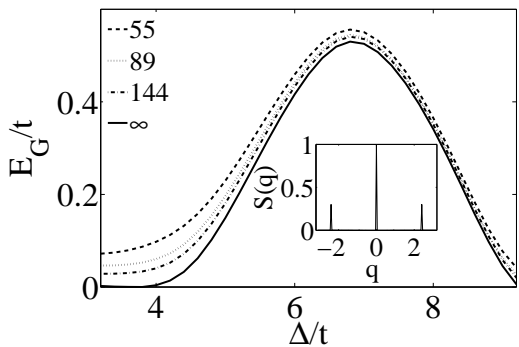}
\vspace*{-0.3cm}
\caption{Excitation gap of the ICDW as a function of $\Delta$ for the quasi-disordered case with $U=5t$ and $V=0$, and different lattice sizes;  
(inset) Static structure factor $S(q)$ for $\Delta=7t$~($N=144$); $S(q)$ 
presents side-peaks corresponding to the $1/(1-\beta)$ quasi-periodicity.}
 \vspace*{-0.3cm}
\label{fig:3}
\end{figure}

\begin{figure}[t]
\includegraphics[clip=true,width=0.8\columnwidth]{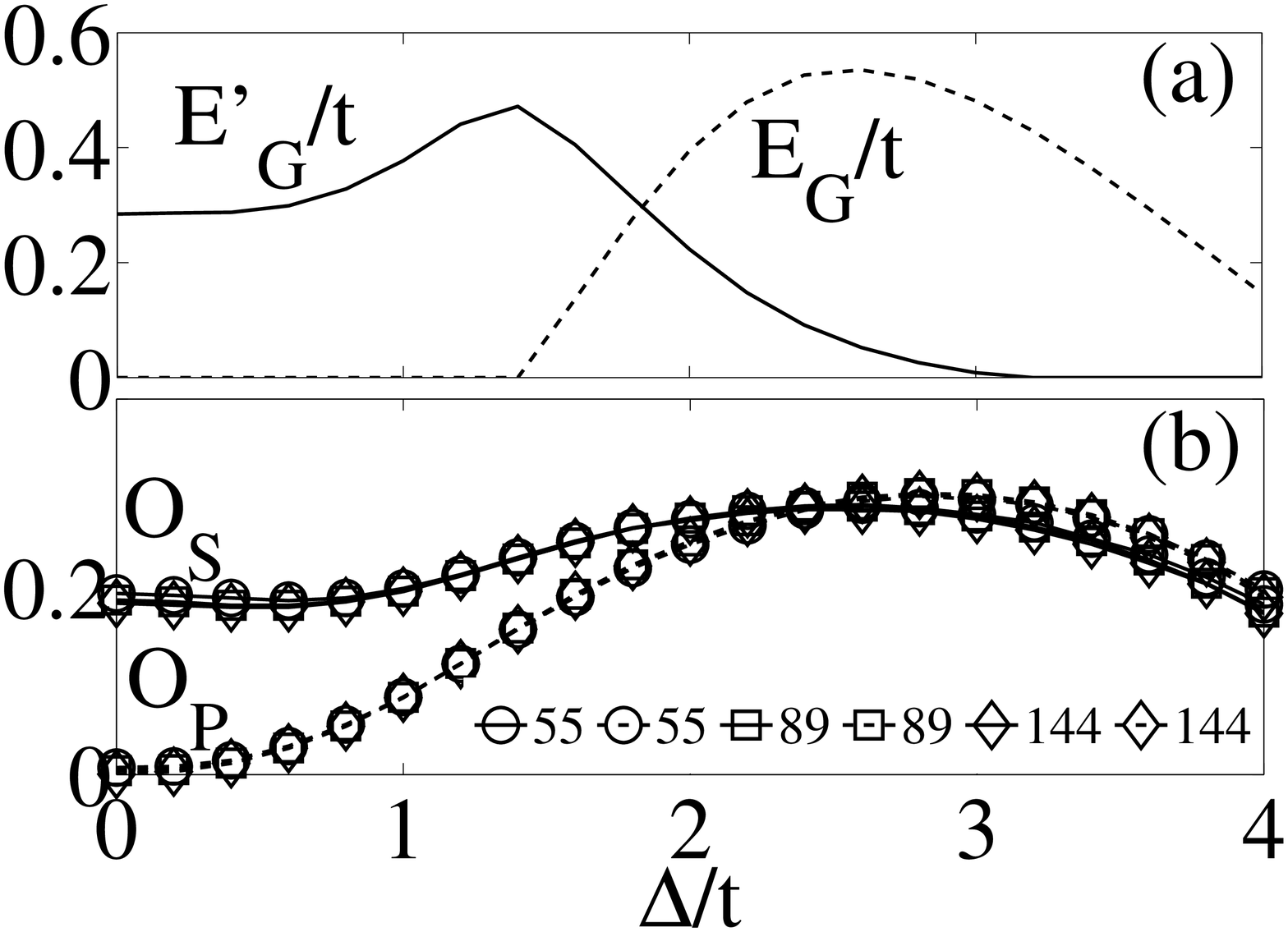}
\vspace*{-0.3cm}
\caption{(a) Gaps $E_G$ and $E'_G$~(extrapolated to infinite-size systems from our results for $N=55$, $89$ and $144$) 
as a function of $\Delta$ for the quasi-disordered case with $U=5t$, and $V=3.3t$; 
(b) For the same parameters, ${\cal O}_S$ and ${\cal O}_P$, for different lattice sites.}
 \vspace*{-0.3cm}
\label{fig:4}
\end{figure}

For $V_c^{(1)}<V<V_c^{(2)}$ we obtain that a gapped phase survives up to a critical $\Delta$ given approximately by the Haldane-like gap at $\Delta=0$~(Figs.~\ref{fig:1}(c) and~\ref{fig:2}(a)),  the HI-BG transition being identified by the closure of the energy gap.   This phase presents a finite ${\cal O}_S$. The survival of this gapped phase for finite uniform disorder 
resembles the persistence of the Haldane phase for weak random magnetic fields in spin-$1$ chains~\cite{Brunel1998,Orignac1998,Nishiyama1998}. This phase however is characterized by a finite ${\cal O}_P$ 
which grows with increasing $\Delta$~(Fig.~\ref{fig:2}(b)). Whereas in the pure HI phase ${\cal O}_P=0$ 
due to the fluid-like spatial distribution of defects $\delta n=\pm 1$, the finite ${\cal O}_P$ relates to the 
growing localization of defects with increasing disorder, while still keeping string order. 
Resorting to the language of ``solid-on-solid'' models~\cite{DenNijs1989}, we may map defects 
$\delta n=\pm 1$ into spin-$1/2$ particles, and sites with $\delta n=0$ as empty sites. In this analogy the pure HI phase 
would be an AF-spin-ordered fluid, whereas the DW phase would be an AF-spin-ordered solid. The Haldane-like phase in the presence 
of disorder could be understood as an AF-spin-ordered glass-like in this analogy.
 
On the contrary the DW is destroyed even for very weak disorder, similar to the destruction of the AF phase in spin-$1$ chains~\cite{Brunel1998,Orignac1998}. 
This results from the proliferation of domain walls in accordance with the Imry-Ma argument~\cite{Imry1975}.  
Our results show for $V>V_c^{(2)}$ a large reduction of the gap for $\Delta\ll V$, 
as well as the expected domain formation. However, 
the convergence of the gap towards zero with the system size is very slow, since the length of the 
Imry-Ma domains scales as $1/\sqrt{\Delta}$, and hence much larger lattice sizes are demanded 
to observe the gap destruction at very weak disorder.

We consider at this point the case of quasi-disordered potentials. Such potentials may be easily generated in ultra-cold gases by means of bichromatic lattices~\cite{Roati2008,Lye2007}, in which a second lattice incommensurate with the first lattice is added. In that case the system experiences an on-site energy $\epsilon_j=\Delta\cos (2\pi\beta j+\phi)$, in which $\beta$ characterizes the incommesurability of both lattices, and $\Delta$ is given by the potential depth of the second lattice. In the following we consider $\beta=(\sqrt{5}-1)/2$. The chain length is chosen following the Fibonacci series $N=55,89,144,233$ in order that the potential is closest to be periodic with $N$. We have checked that in most cases the results are not affected by the choice of $\phi$, and we hence set $\phi=0$ in the following.

Quasi-periodic potentials allow for an ICDW~\cite{Roscilde2008,Roux2008}, which is a localized but gapped phase, contrary to the BG. The ICDW may be understood from the beating of both lattice frequencies which leads to super-wells with a characteristic length scale $1/(1-\beta)$. A filling $1-\beta$ (or $\beta$) hence leads to the formation of an approximate density wave. For filling factor $n=1$, a possibly gapped localized phase (i.e. a generalized ICDW) may occur as well, and a MI-SF-ICDW-BG scenario cannot be excluded~\cite{Roux2008}. Our results for $V=0$ show a clearly gapped localized ICDW
whose gap converges to a finite value when extrapolating to infinite size~(Fig.~\ref{fig:3}).  This phase is characterized by a peak in the static structure factor $S(k)$ given by the approximate $1/(1-\beta)$ periodicity~(inset of Fig.~\ref{fig:3}).
Hence, at $V=0$ we observe a MI-SF-ICDW-BG phase diagram~(Fig.~\ref{fig:1}(d))~\cite{footnote-ICDW-BG}. The ICDW-BG transition is characterized by  an algebraic decay of $G(r)$ with Luttinger exponent  $K\simeq 1$, compatible with results obtained for spinless fermions~\cite{Vidal1999}. At the SF-MI transition we obtain that $K$ decreases with growing $V$.  For $V<V_c^{(1)}$ the MI-SF-ICDW-BG topology is preserved,  but the MI and SF regions shrink vanishing at $V_c^{(1)}$. 
For $V_c^{(1)}<V<V_c^{(2)}$ and low $\Delta$, we obtain as for the uniform disorder case a HI phase with growing ${\cal O}_P$ 
when $\Delta$ increases~(Fig.~\ref{fig:4}~\cite{footnote1}). This is again due to the growing localization of defects 
$\delta n=\pm 1$ by the quasi-disorder while keeping string-order. 
However, contrary to the uniform disorder case, the quasi-disordered case allows for the generalized 
ICDW phase which has also finite ${\cal O}_S$ and ${\cal O}_P$. As a consequence, the increasing pinning of defects for increasing $\Delta$ leads to a connection between HI~($E'_G >0$) and generalized ICDW~($E_G>0$), without any gapless region in between~(Fig.~\ref{fig:4}). Our DMRG calculations, including the determination of fidelity susceptibility~\cite{You2007}, show that the generalized ICDW and the HI phases form an overall gapped region. 
Finally, for $V>V_c^{(2)}$ we observe the opening of a finite DW region, which contrary to the uniform disorder case is not immediately destroyed at small $\Delta$. The DW region undergoes at finite $\Delta$ an Ising transition into the generalized ICDW phase, the transition being identified by an algebraic decay of $G(r)$.

In summary, uniform disorder and quasi-disorder lead to different phase diagrams for polar bosons in
optical lattices. For uniform disorder a HI phase with finite parity survives up to a finite disorder where a HI-BG transition occurs,
whereas the DW phase becomes a BG for any finite disorder. On the contrary, in the presence of quasi-disorder the HI phase continuously
connects with a generalized ICDW phase without any intermediate critical region, and the DW survives up to a finite disorder. 
The predicted phases may be detected using state of the art techniques. In particular, the string order may be
detected using similar site-resolved techniques as those recently used for detecting non-local parity order~\cite{Endres2011}.  

\acknowledgments
We thank T. Vekua for useful discussions. X.~D. and L.~S. are supported by the DFG~(SA1031/6), the German-Israeli Foundation, and the Cluster of
Excellence QUEST. E.~O. and A.~M. acknowledge support from the CNRS PEPS-PTI project "Strong correlations
and disorder in ultracold quantum gases", and A.~M. from the Handy-Q ERC project.

\newpage


\begin{thebibliography}{10}
\bibitem{footnote-ICDW-BG} Our numerical calculations cannot resolve whether the 
ICDW-BG boundary is a critical line or rather a very narrow SF region.
\bibitem{footnote1} For quasi-disorder we employ the definition of ${\cal O}_{S,P}$ in the text, averaging 
over initial sites $i$.
\end{thebibliography}
\end{document}


\title{Supplementary material for ``Polar bosons in one-dimensional disordered optical lattices''}

\author{Xiaolong Deng}
\affiliation{Institut f\"ur Theoretische Physik, Leibniz Universit\"at Hannover, Appelstr. 2, D-30167 Hannover, Germany}
\author{Roberta Citro}
\affiliation{Dipartimento di Fisica ``E. R. Caianiello'' and Spin-CNR, Universit\`a degli Studi di Salerno, Salerno, Italy}
\author{Edmond Orignac}
\affiliation{Laboratoire de Physique de l'\'Ecole Normale Sup\'erieure de Lyon, CNRS-UMR5672, 69364 Lyon Cedex 7, France}
\author{Anna Minguzzi}
\affiliation{Universit\'e Grenoble I and CNRS, Laboratoire de Physique et Mod\'elisation,
des Milieux Condens\'es UMR 5493, Maison des Magist\`eres, B.P. 166, 38042 Grenoble, France}
\author{Luis Santos}
\affiliation{Institut f\"ur Theoretische Physik, Leibniz Universit\"at Hannover, Appelstr. 2, D-30167 Hannover, Germany}

\maketitle

\section{Error bars for the model with uniform disorder}
\label{sec:Errorbars}

\begin{figure}[ht]
\includegraphics[clip=true,width=0.75\columnwidth]{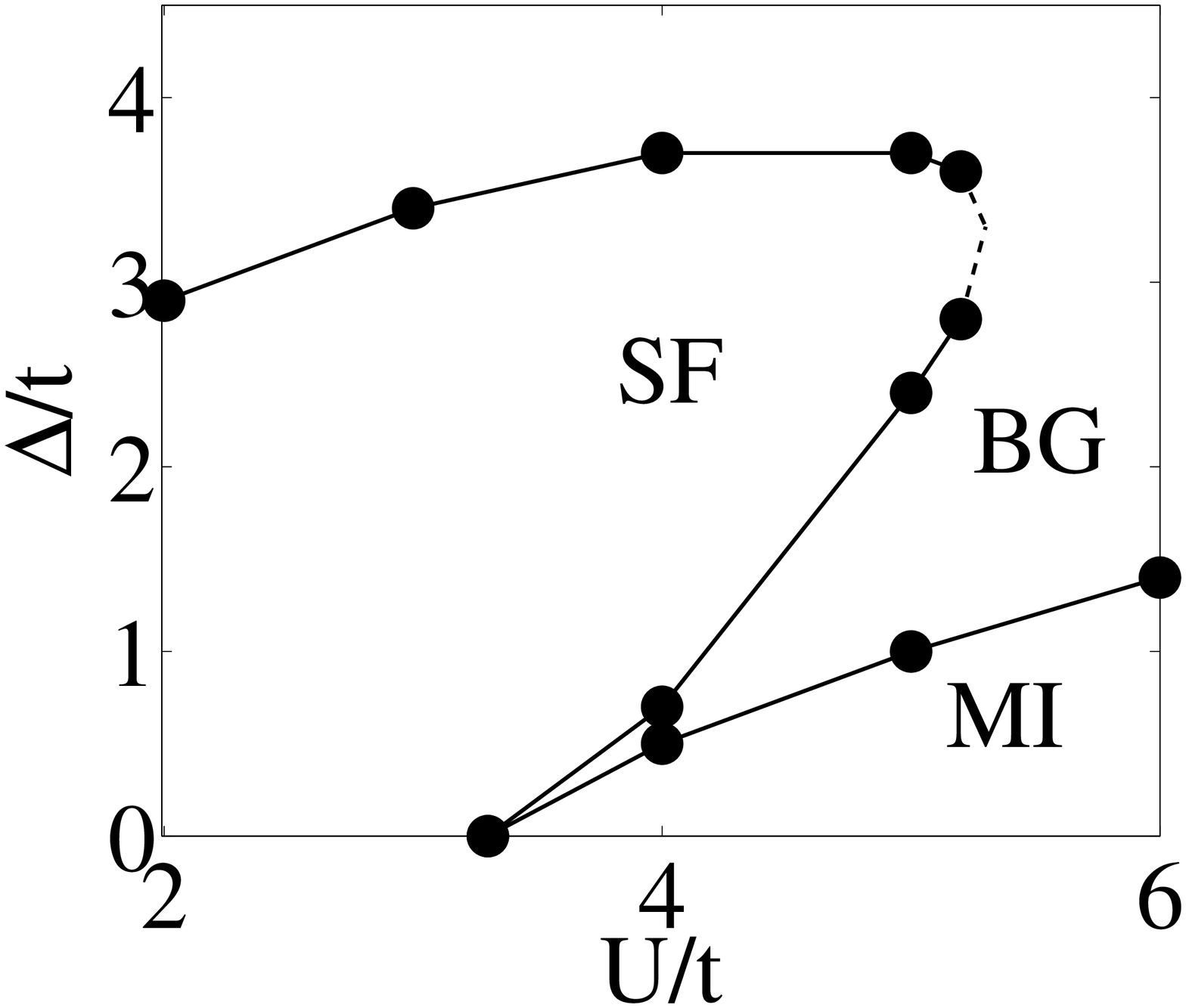}
\caption{Phase diagram for the non-polar Bose-Hubbard model with uniform box disorder, for a chain length $L=89$, with $30$ disorder realizations per point.}
\label{fig:1SM}
\end{figure}

We determine the SF-BG boundary by monitoring the behavior of the single-particle correlation $G(r)$, which has a power-law decay in the SF phase. This decay can be fitted very well using conformal field corrections for open boundary conditions~\cite{Roux2008}, allowing for the calculation of the corresponding Luttinger parameter $K$. The SF-BG phase boundary is determined at the point in which $K=3/2$~\cite{Giamarchi.Schultz.1988}. In order to benchmark our calculations we have established the phase diagram for disorded non-polar gases. Figure~\ref{fig:1SM} shows that the calculated SF-BG boundary is in excellent quantitative agreement with that obtained using quantum Monte Carlo methods~\cite{Prokofev.Svistunov.1998}. 

In order to determine the error bars in Fig.~1c of the main text, we have studied for different $V$ and $\Delta$ values up to $60$ disorder realizations. We randomly pick different sub-sets of these realizations and average for each sub-set the correlation $G(r)$. The error bars indicate for a given $V$ the regime of $\Delta$ values 
in which for some sub-sets, but not for all, correlations are best fitted by a power-law with $K=3/2$.
We establish the error bars between gapped and gapless phases in Fig.~1c of the main text, i.e. MI-BG and HI-BG, using a similar procedure. We evaluate again a large number of disorder realizations for different lattice sizes. We perform finite-size scaling to obtain the excitation gap. The error-bar region is that at which some disorder realizations present finite gap whereas other realizations are not gapped.

\begin{figure}[t]
\includegraphics[clip=true,width=0.8\columnwidth]{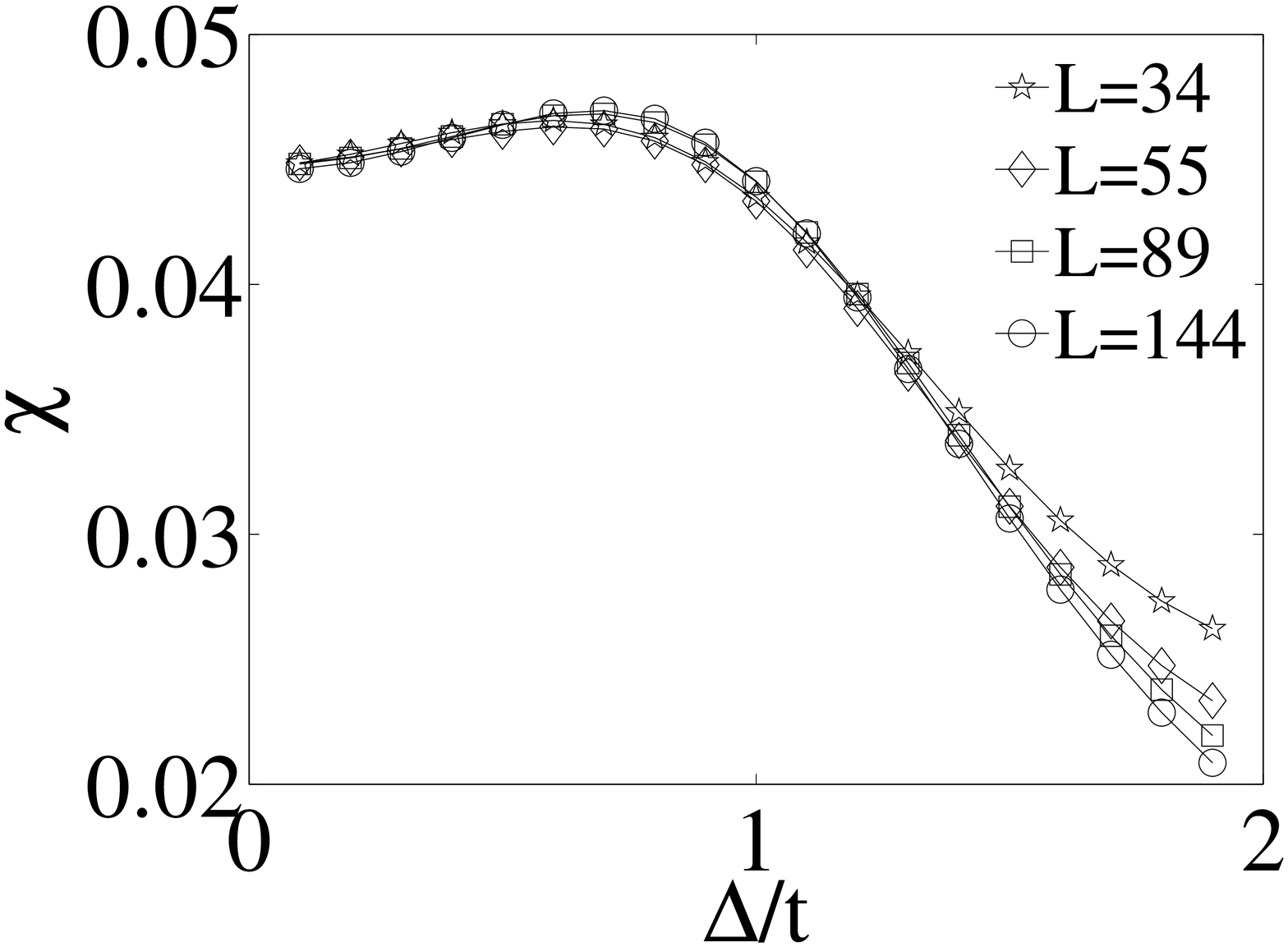}
\caption{Fidelity susceptibility as a function of disorder in the HI phase for the quasi-periodic case with OBC. The parameters are $U/t=5$, $V/t=3.1$ and the initial angle of the quasi-periodic potential is $\phi=0$.}
\label{fig:2SM}
\end{figure}

\section{Fidelity susceptibility and energy gap within the HI-ICDW phase}
\label{sec:Fidelity}

\begin{figure*}[t]
\includegraphics[clip=true,width=0.8\columnwidth]{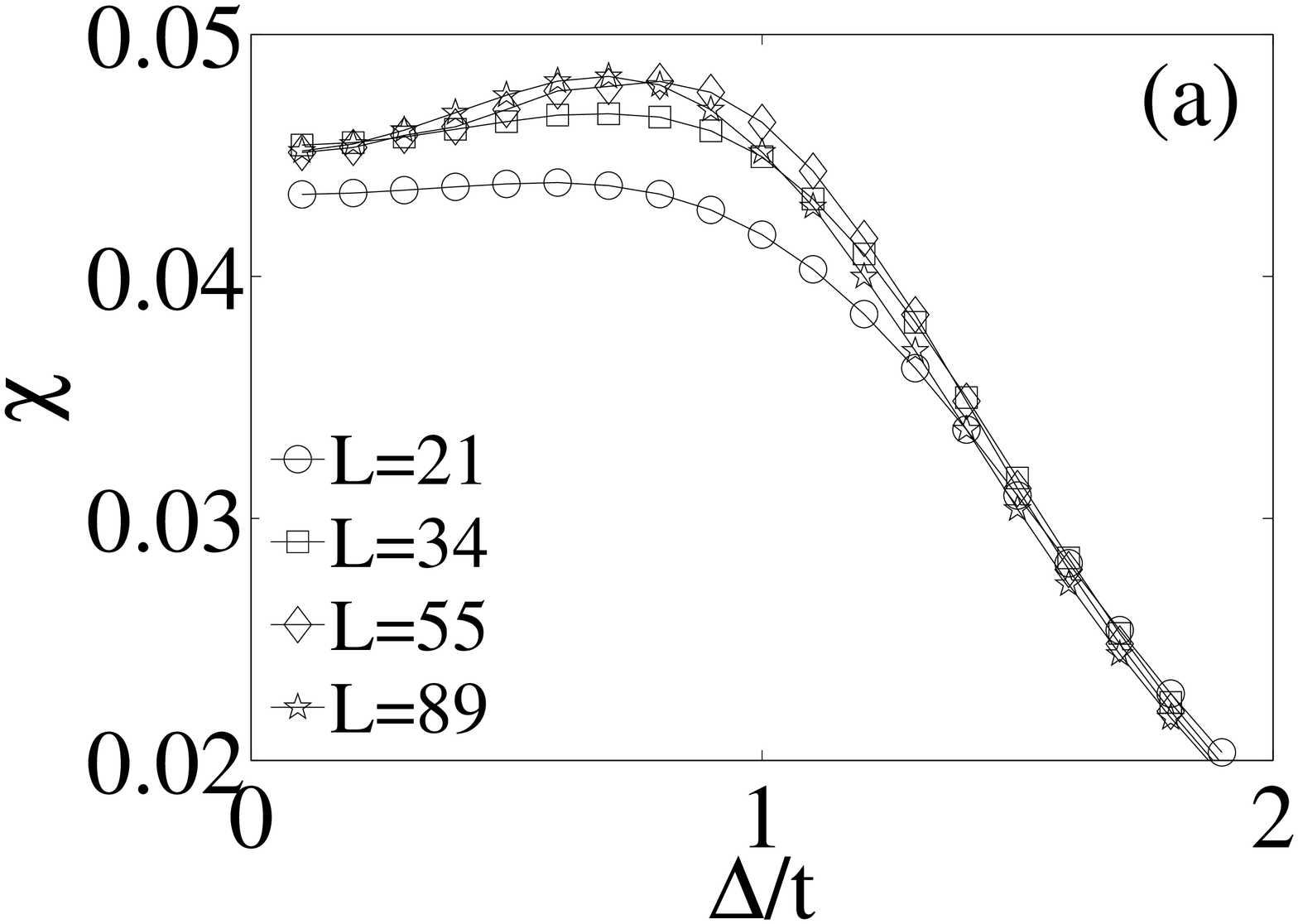}
\includegraphics[clip=true,width=0.8\columnwidth]{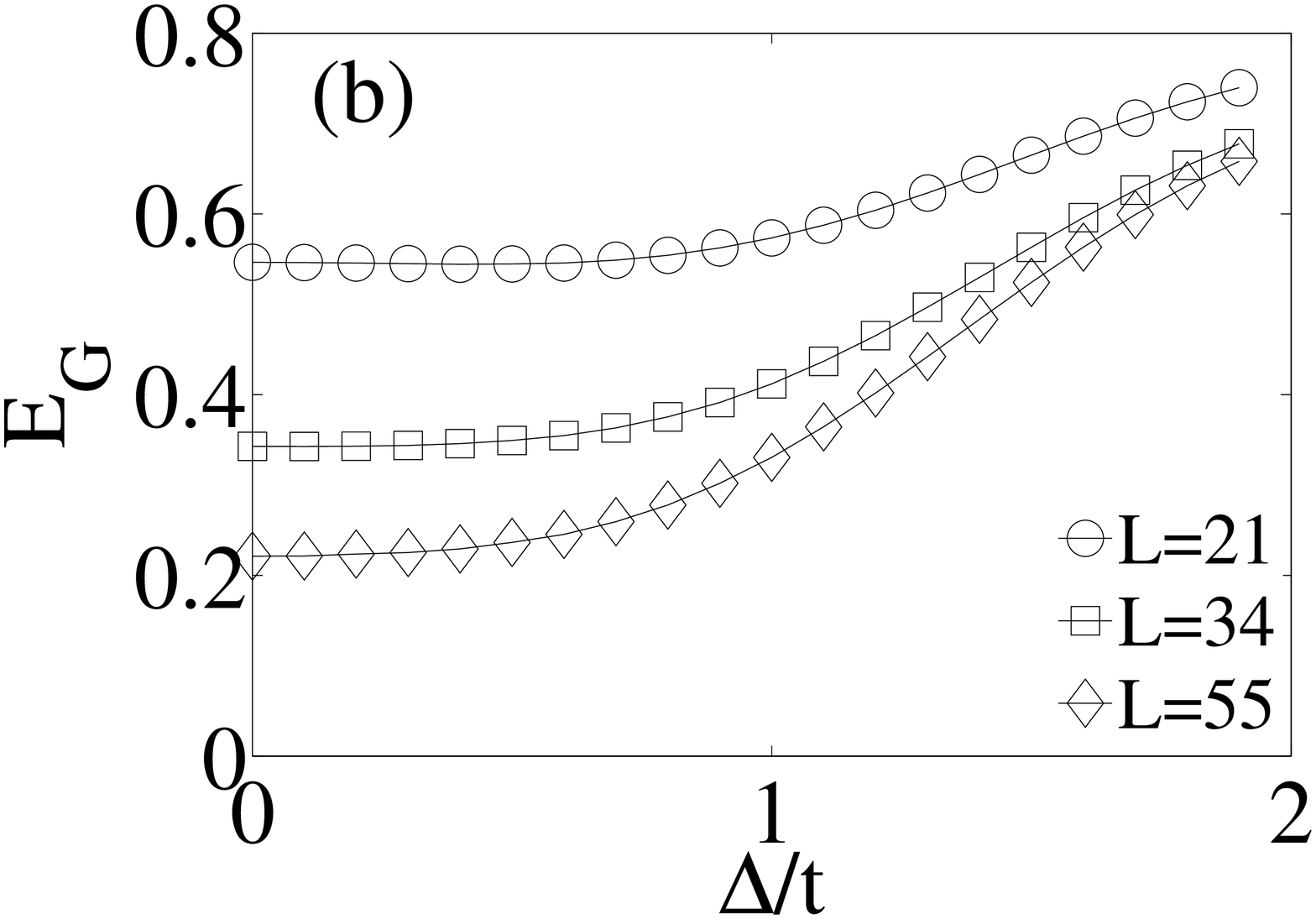}
\caption{Fidelity susceptibility~(a) and energy gap~(b) as a function of disorder for the case of Fig.~\ref{fig:2SM} but with PBC.}
\label{fig:3SM}
\end{figure*}

\begin{figure*}
\includegraphics[clip=true,width=0.8\columnwidth]{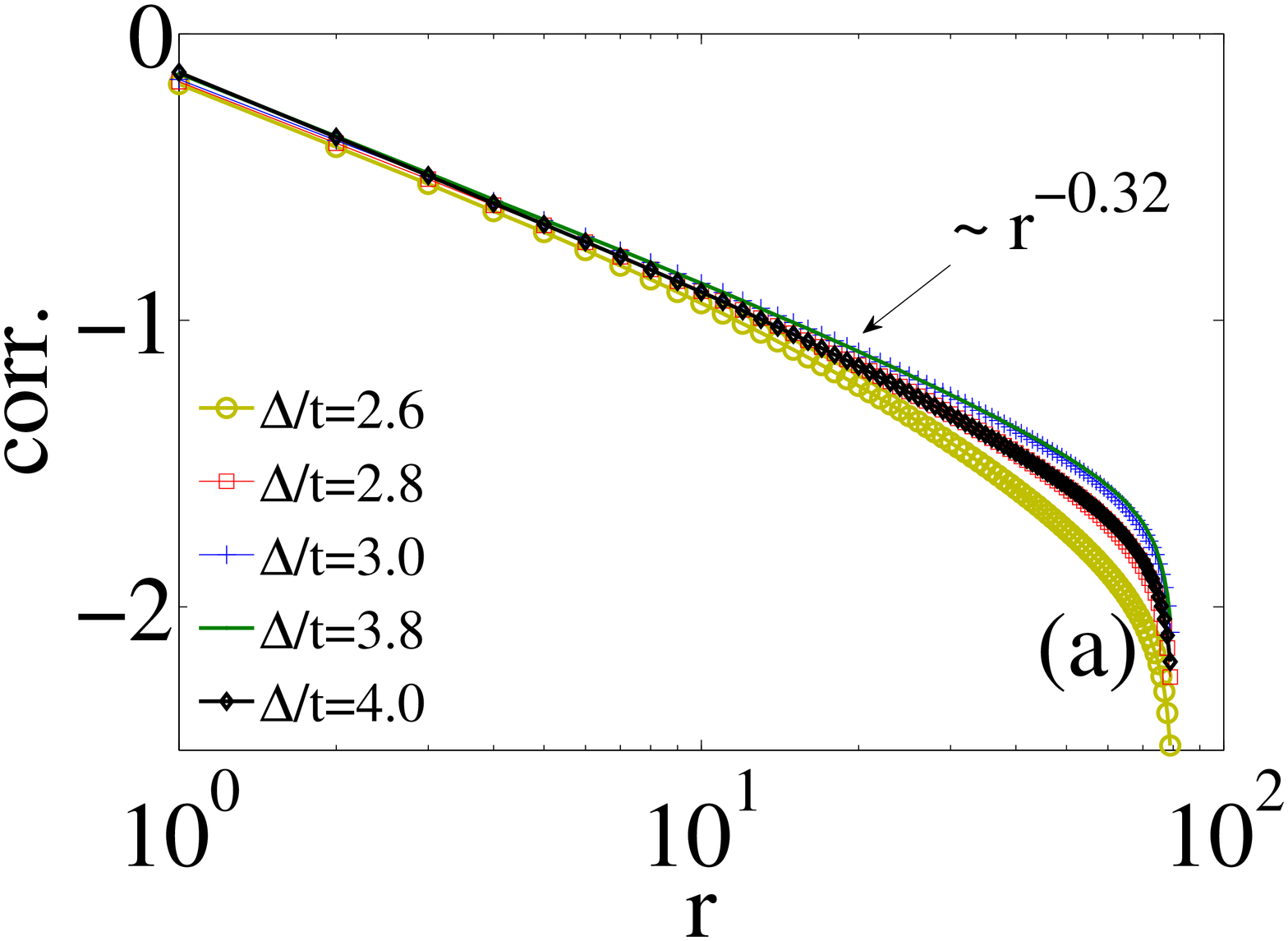}
\includegraphics[clip=true,width=0.8\columnwidth]{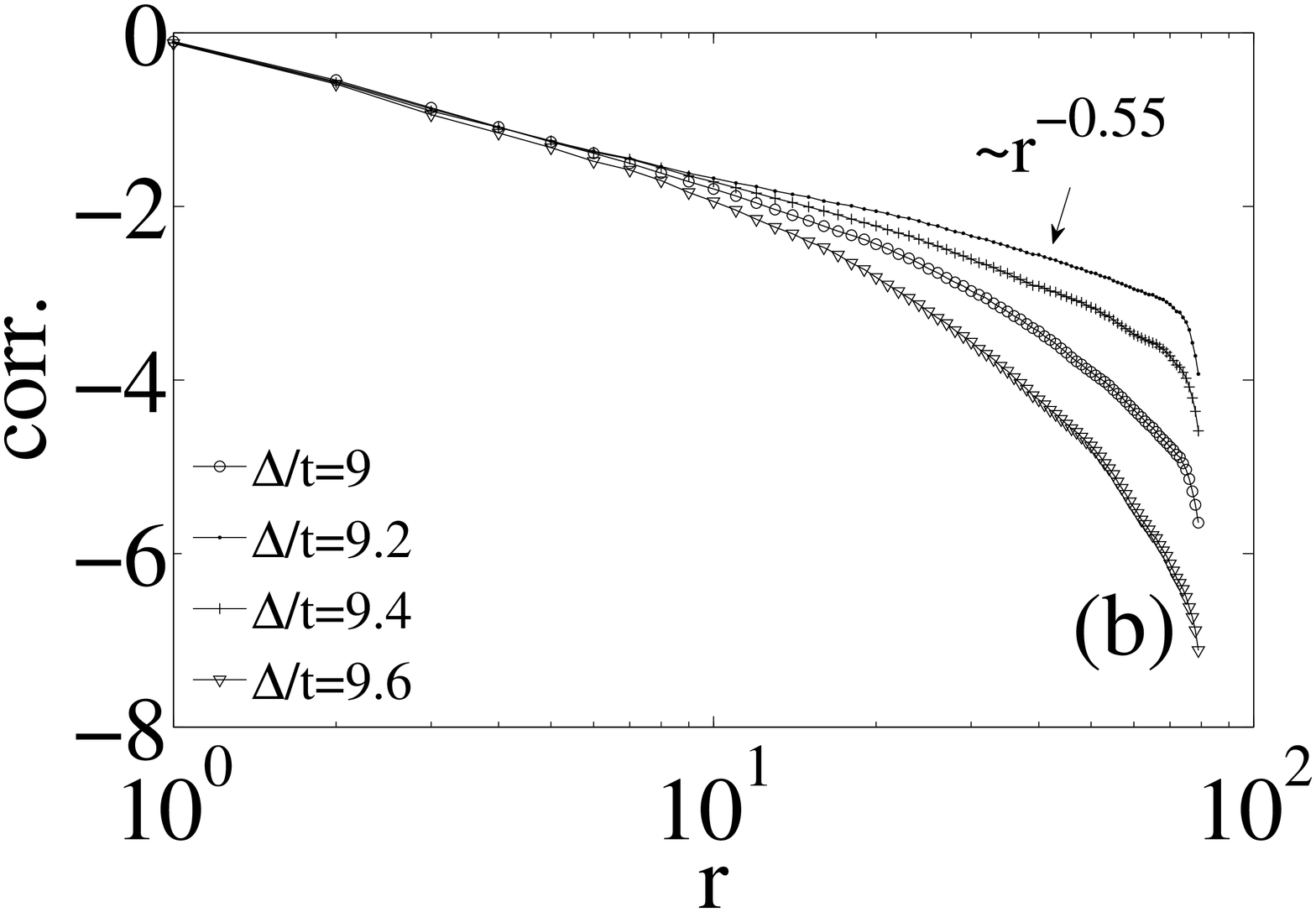}
\caption{Single-particle correlation in logarithmic scale as a function of distance, 
under different quasi-disorder amplititudes for $L=89$, $U/t=5$ and $V/t=0$. 
(left) Results at the MI-SF-ICDW region. At the SF-ICDW region we 
find $K\simeq 1.5$. (right) Results in the vicinity of the ICDW-BG transition. At the 
transition at $\Delta/t \approx 9.2$, the correlation decays algebraically, 
with a Luttinger exponent $K=1$. 
In both graphs the correlations have been averaged over $20$ different initial angles $\phi\in [0,2\pi)$.}
\label{fig:4SM}
\end{figure*}

As commented in our paper, we find that for the quasi-periodic case the HI phase and the generalized ICDW phase are adiabatically connected. We provide in this section further details supporting this conclusion.
The fidelity susceptbility for the ground state as a function of disorder is defined as~\cite{You2007}
\begin{eqnarray}
\chi = \frac{2\left[1-\left|\langle \psi(\Delta)|\psi(\Delta+{\rm d}\Delta)\rangle\right|\right]}{L({\rm d}\Delta)^2},
\end{eqnarray}
where $\psi(\Delta)$ is the ground state for a given disorder $\Delta$ (and fixed $U$, $V$, $t$), and 
$d\Delta$ is a small variation of $\Delta$. 
In general, the fidelity susceptibility shows a marked peak at a phase transition. Fig.~\ref{fig:2SM} 
shows our results for the fidelity susceptibility in the HI phase for the case of open boundary 
conditions~(after properly eliminating edge states). Our results for various chain sizes show no peak feature of the fidelity susceptibility. Similar results are found using periodic boundary conditions~(PBC) for up to $L=89$ sites, as shown in Fig.~\ref{fig:3SM}(a). 

In Fig. 4(a) of our paper we showed that there is always a significant gap~(either $E_G$ or $E_G'$) and hence that no gapless region occurs between HI and ICDW. We have analyzed as well the excitation gap for the case of PBC. Note that in this case there are naturally no edge states, and hence we may study a single gap, instead of considering the gaps $E_G$ and $E_G'$ as for the calculations with open boundary conditions~(OBC). We have paid particular attention to the region in Fig.4 (a) of our paper in which the system has a crossover between having a dominant $E_G'$ gap into a dominant $E_G$ gap. As depicted in Fig. \ref{fig:3SM}(b) in that region the PBC results for the gap show no trace of a gapless region.

\section{ICDW-BG and MI-SF transitions}
\label{sec:Transitions}
For the determination of the properties of the quasi-periodic system at the ICDW-BG and MI-SF boundaries, we average our results for the single-particle correlations over up to $20$ quasi-periodic potentials with different $\phi\in [0,2\pi)$, and fit their decay by an exponential or a power-law curve. The MI-SF transition is provided by the transition from an exponential to power-law decay. The critical Luttinger exponent $K$ along the boundary of MI-SF changes with increasing nearest-neighbor interactions $V$~(the case at $V=0$ is depicted in Fig.~\ref{fig:4SM}(a)). We obtain as well a gapless critical ICDW-BG line, which is characterized by a power-law decay with constant $K \approx 1$~(Fig.~\ref{fig:4SM} (b)).

\newpage